# Magnon-Polaron Driven Thermal Hall Effect in a Heisenberg-Kitaev Antiferromagnet

N. Li,[1,*] R. R. Neumann,[2,*] S. K. Guang,[1,*] Q. Huang,[3] J. Liu,[3] K. Xia,[1] X. Y. Yue,[4] Y. Sun,[4] Y. Y. Wang,[4] Q. J. Li,[5] Y. Jiang,[5] J. Fang,[6] Z. Jiang,[6] X. Zhao,[7] A. Mook,[8] J. Henk,[2] I. Mertig,[2,†] H. D. Zhou,[3,†] and X. F. Sun[1,4,†]

[1] Department of Physics, Hefei National Laboratory for Physical Sciences at Microscale, and Key Laboratory of Strongly-Coupled Quantum Matter Physics (CAS), University of Science and Technology of China, Hefei, Anhui 230026, People's Republic of China

[2] Institute of Physics, Martin Luther University Halle-Wittenberg, Halle (Saale), Germany

[3] Department of Physics and Astronomy, University of Tennessee, Knoxville, Tennessee 37996-1200, USA

[4] Institute of Physical Science and Information Technology, Anhui University, Hefei, Anhui 230601, People's Republic of China

[5] School of Physics and Optoelectronics, Anhui University, Hefei, Anhui 230601, People's Republic of China

[6] School of Physics, Georgia Institute of Technology, Atlanta, GA 30332, USA

[7] School of Physical Sciences, University of Science and Technology of China, Hefei, Anhui 230026, People's Republic of China

[8] Institute of Physics, Johannes Gutenberg University, Mainz, Germany

[*] These authors contributed equally: N. Li, R. R. Neumann, S. K. Guang.

[†] Email: ingrid.mertig@physik.uni-halle.de; hzhou10@utk.edu; xfsun@ustc.edu.cn

**The thermal Hall effect, defined as a heat current response transversal to an applied temperature gradient, is a central experimental probe of exotic electrically insulating phases of matter. A key question is how the interplay between magnetic and structural degrees of freedom gives rise to a nonzero thermal Hall conductivity (THC). Here, we present evidence for an intrinsic thermal Hall effect in the Heisenberg-Kitaev antiferromagnet and spin-liquid candidate $Na_2Co_2TeO_6$ brought about by the quantum-geometric Berry curvature of so-called magnon polarons, resulting from magnon-phonon hybridization. At low temperatures, our field- and temperature-dependent measurements show a negative THC for magnetic fields below 10 T and a sign change to positive THC above. Theoretically, the sign and the order of magnitude of the THC cannot be solely explained with magnetic excitations. We demonstrate that, by incorporating spin-lattice coupling into our theoretical calculations, the Berry curvature of magnon polarons**

**counteracts the purely magnonic contribution, reverses the overall sign of the THC, and increases its magnitude, which significantly improves agreement with experimental data. Our work highlights the crucial role of spin-lattice coupling in the thermal Hall effect.**

Topological phases of matter have received enormous attention in solid state physics not only for their exceptional fundamental properties, but also for their potential technological impact. For example, topological band insulators feature protected, dissipationless edge channels[1–3], and topological order in strongly correlated electron systems (e.g., quantum spin liquids[4–6]) may be a route to fault-tolerant quantum computing.[7,8] Harvesting the potential of these exotic phases requires a reliable technique for their detection and characterization. An important probe for topological phases in insulators is the thermal Hall effect (THE), which denotes a transverse heat current response to a longitudinal temperature gradient.[9,10] Its *intrinsic* contribution is an invaluable probe of the Berry curvature, that is, a quantum-geometric property acting on the inherent quasiparticles, e.g., Majorana fermions[11,12], triplons[13,14], photons[15,16], and magnons[17–23], like a fictitious magnetic field. However, the ubiquitous phonons (quanta of lattice vibrations) interact and potentially hybridize with the aforementioned quasiparticles due to spin-lattice coupling (SLC).[24–26] The band inversions of these quasiparticle-phonon hybrids establish another source of Berry curvature that may even dominate the low-temperature THE because of the low acoustic phonon energies. Hence, detailed understanding of SLC and its effects on intrinsic heat transport is required.

In this joint experimental and theoretical work, we report the thermal Hall conductivity (THC) $\kappa_{xy}$ of the Kitaev spin-liquid candidate $Na_2Co_2TeO_6$ (NCTO), which has attracted considerable attention recently.[27–35] Our field- and temperature-dependent measurements reveal a negative THC for out-of-plane magnetic fields below 10 T and a positive THC above 10 T at low temperatures. We attribute this sign change to a field-driven magnetic phase transition. As we demonstrate theoretically, magnons fail to explain not only the overall sign of the THC, but also its order of magnitude, as the THC is underestimated by a factor of 10. By taking SLC into account, magnons and phonons form hybrid quasiparticles, i.e., magnon polarons. The Berry curvature at the resulting avoided crossing between the lowest magnon and the acoustic phonon band is of opposite sign compared to the low-energy magnon Berry curvature without SLC. Hence, we reproduce both the correct overall sign and the order of magnitude of the experimental THC. The sign reversal of the THC due to the hybridization of phonons and magnons is one of our main findings that is visualized in Figure 1. Our results indicate the pivotal role of the SLC in thermal transport, which may be also relevant to the interpretation of the THC in related Heisenberg-Kitaev magnets.[36–38]

NCTO is composed of $Co^{2+}$ ions, arranged in layers of honeycomb lattices, whose effective $S = ½$ spins order in antiferromagnetic (AFM) zigzag chains (cf. Figure 2a). Employing a three-thermometer setup (panel b; cf. Methods), we have measured the temperature dependence of the longitudinal thermal conductivity $\kappa_{xx}(T)$ of NCTO at zero magnetic field (panel c). According to previous studies, NCTO enters a magnetically ordered state below the Néel temperature $T_N = 27$ K, followed by two possible spin reorientations around 16 K and 6 K, respectively.[27–30] Our $\kappa_{xx}(T)$ data show no obvious anomalies around 27 K and 16 K, in agreement with reported $\kappa_{xx}(T)$ data,[39] but a slope change below around 6 K possibly related to a spin reorientation.[28,29] At sub-Kelvin temperatures, $\kappa_{xx}(T)$ roughly follows a $T^{1.2}$ behavior, which is at variance with the $T^3$ or $T^2$ behavior expected for the phonon thermal conductivity at low temperatures in 3 or 2 dimensions, respectively. Since magnons are frozen out at temperatures corresponding to energies below the spin-wave gap, their contribution does not explain the observed scaling law either. Therefore, this $T^{1.2}$ behavior may indicate the significance of interactions between phonons and magnons.

The field dependence of $\kappa_{xx}(B)$ measured at various temperatures with ***B*** ∥ ***c*** is depicted in Figures 2d and 2e. At $T < 1.56$ K, $\kappa_{xx}(B)$ decreases quickly with increasing field to reach a minimum around 4 T, and then shows a weak field dependence up to 14 T. At 2.2 K, 2.7 K, and 3.2 K, $\kappa_{xx}(B)$ manifests a double-valley structure with valleys around 2 T and 10 T, respectively. At even higher temperatures, $\kappa_{xx}(B)$ exhibits a broad valley in the range of 5 – 10 T. Similar observations have been reported in Ref. 40.

Apart from the complex longitudinal thermal conductivity, we find a peculiar field dependence of the THC $\kappa_{xy}(B)$, which we have measured at various temperatures below $T_N$ (cf. Figures 3a,b). At $T \leq 2.2$ K, with increasing field, $\kappa_{xy}(B)$ first exhibits a negative Hall response reaching a minimum around 3 to 5 T; then it changes to positive sign around 10 T and increases at higher fields. At 3.2 K and 5.4 K, $\kappa_{xy}(B)$ curves show a positive peak at low fields, followed by two zero crossings with increasing field. At 7.8 K, $\kappa_{xy}(B)$ is positive without sign reversal. We have plotted the temperature dependence of $\kappa_{xy}/T$ at several fields in Figure 3c. It is evident that at $B = 3$ T and 5 T, with increasing temperature, $\kappa_{xy}$ is negative and reaches a minimum around 2 K, and then changes to positive sign around 3 K to 4 K. The thermal Hall angle $\kappa_{xy}/\kappa_{xx}$ possesses a minimum around 4 T and changes to positive sign around 10 T at temperatures below 2.2 K (cf. Figure 3d). The largest absolute value of $\kappa_{xy}/\kappa_{xx}$ is around 2 % at 0.78 K and 4 T.

In a magnetic insulator, the observed $\kappa_{xy}$ may have several origins, including phonons, magnons, and fractionalized exotic quasiparticles such as spinons. In experiments on nonmagnetic insulators, $\kappa_{xy}$ of phonons does not exhibit a sign change.[40,41] For spinons, a nonzero $\kappa_{xy}$ has only been observed in a quantum spin liquid with disordered spins.[42–44] Apparently, these two scenarios cannot explain the sign-reversible and non-monotonic $\kappa_{xy}(B)$ observed in the AFM state of NCTO. Moreover, the 2 % thermal Hall angle is exceptionally large for an insulator. The expected value, either originating from phonons or magnons, is typically around 0.3 % to 0.6 % or even lower.[45] To our knowledge, only the pyrochlore magnet $Yb_2Ti_2O_7$ shows a thermal Hall angle of 2 % in its quantum spin-liquid state.[44]

The experimental results on the transverse transport properties of NCTO are subsequently explained by an effective, semi-quantitative model. The starting point is the Heisenberg-Kitaev-Gamma-Gamma' (HKGG') Hamiltonian that encompasses Heisenberg exchange [$J_r$ ($r = 1, 2, 3$)] up to 3rd nearest neighbors and Kitaev ($K$), Gamma ($\Gamma$), Gamma' ($\Gamma'$) interactions between nearest neighbors. The magnetic field $\boldsymbol{B}$ enters via the Zeeman Hamiltonian (cf. Methods). Here, we are interested in out-of-plane fields $\boldsymbol{B} \parallel \boldsymbol{c}$. Several parameter sets of the spin Hamiltonian have been determined for NCTO [cf. Supplementary Information (SI)]. In the following, we choose $J_1 = -3.2$ meV, $J_2 = 0.1$ meV, $J_3 = 1.2$ meV, $K = 2.7$ meV, $\Gamma = -2.9$ meV, and $\Gamma' = 1.6$ meV.[31] This parameter set (referred to as tc+) reproduces the critical fields in experimental reports on field-induced magnetic phase transitions (cf. SI) and, as presented below, provides the best agreement with the experimental THC. Results for other parameter sets are reported in the SI. The weak inter-layer coupling is neglected.

The antiferromagnetic ground state of the Hamiltonian at zero field is characterized by zigzag chains with intra-chain ferromagnetic and inter-chain antiferromagnetic order. Applying a magnetic field cants the spins slightly, but they remain confined to the $bc$ plane (cf. left inset in Figure 4a). At the critical field of $B_{c1} = 10.8$ T the system passes a first-order phase transition into a spin-flop state, in which the spins lie within the $ac$ plane (ferromagnetic component along $c$ and Néel vector along $a$; cf. right inset in Figure 4a). The magnetization saturates at $B_{c2} = 31.2$ T, at which the fully field-polarized phase is reached. These critical fields are supported by magnetometry measurements (cf. SI).

The diagonalization of the linearized Hamiltonian yields the four magnon bands $\varepsilon_{nk}$ ($n = 1, 2, 3, 4$). Because of the spin-½ nature of the local magnetic moments, significant quantum fluctuations are expected, which are captured by an effective rescaling of the magnon energies $\varepsilon_{nk} \to \Lambda \varepsilon_{nk}$. The intrinsic contribution to the THC is computed with the linear response formalism (cf. Methods).

Figure 4a shows $\kappa_{xy}$ versus $B_z$ as computed from free-magnon calculations for six temperatures. In the low-field phase, $\kappa_{xy}$ is positive and changes sign at $B_{c1}$ for all temperatures. This sign change is thus linked to the magnetic phase transition. However, the overall sign of $\kappa_{xy}$ is at variance with the measured data (cf. Figure 3a). Attributed to the first-order transition, $\kappa_{xy}$ is discontinuous at the phase transition, with maximum left and minimum right of $B_{c1}$. Moreover, the experimental data are underestimated by a factor of 10. Similar calculations with other parameter sets taken from the literature fail to reproduce the data as well (cf. SI).

The foregoing suggests that magnons by themselves are not sufficient to explain the experimental data. We therefore consider a particular SLC arising from spin-orbit coupling (cf. Methods). We neglect vibrations of nonmagnetic ions and other types of SLC for a minimal description. Furthermore, we consider a single acoustic phonon branch in the crystallographic Brillouin zone (that is, two branches in the magnetic Brillouin zone). The relevant energy scale $\lambda$ quantifies the strength of the SLC. The elastic constant $C$ is chosen to yield a phonon velocity of $3000\,\frac{\text{m}}{\text{s}}$, which is supported by heat capacity measurements (cf. SI). We proceed by bosonizing spin and position operators and extend the basis by the two phonon modes. The extended Hamiltonian is then diagonalized, and $\kappa_{xy}$ is computed as before. The SLC strength $\lambda = 0.37$ meV has been fitted as an effective parameter to reproduce the experimental THC at 2.2 K.

Figure 4b displays $\kappa_{xy}$ versus $B_z$ in the presence of SLC. Compared to exclusive magnon transport, the overall sign of $\kappa_{xy}$ is reversed and the sign change at the magnetic phase transition remains intact. Furthermore, $\kappa_{xy}$'s order of magnitude has increased and matches that of the experimental data. In short, agreement with the experiment has increased significantly.

The sign change and the increase of $|\kappa_{xy}|$ are attributed to hybrid quasiparticles that we refer to as magnon polarons. These normal modes are superpositions of magnons and phonons. Their hybrid nature is prominent in the band structure with SLC (Figure 4c) at avoided crossings: their character changes continuously from magnon- (red) to phonon-like (blue). The avoided crossing between the acoustic phonon branch and the lower magnon band generates positive Berry curvature in the lowest band, indicated by a white arrow in panel d. This pronounced low-energy Berry curvature dominates the transport and explains the negative sign in the zigzag antiferromagnetic phase. This finding is contrasted with the Berry curvature in the absence of SLC (panel e). Ignoring the phonon bands (blue in panel c), the magnon bands exhibit a spin-wave gap, and their lowest energies are at the Γ and S points. The Berry curvature of the lowest magnon band (panel e) at these points is negative and positive, respectively. Since the two lower magnon bands are degenerate at S and the upper band exhibits the opposite Berry curvature at S, the Berry curvature at Γ mostly governs the thermal transport at low temperatures. This Berry curvature is, however, opposite to the emerging Berry curvature caused by the hybridization. Thus, there is a competition between pure magnon transport and magnon-polaron transport in the presence of SLC.

The gradual suppression and sign reversal of $\kappa_{xy}$ by the coupling to phonons hold for lower temperatures. At higher temperatures, the magnon bands are strongly populated and the transport coefficient changes sign back (cf. SI). This competitive interplay between phonons and magnons is contrasted by the results of Zhang *et al.* for the honeycomb ferromagnet VI$_3$, which has been modeled with Dzyaloshinskii-Moriya interaction (DMI) as the source of the magnon Berry curvature.[26] There, an amplification of the THC was found due to the SLC. Notably, an attenuation can be found for reversed DMI and, hence, either scenario is within the reach of the DMI model with SLC. In contrast, the HKGG' model with SLC uniquely fixes both the sign of the magnon and the magnon-polaron Berry curvatures and, therefore, their relative sign. This renders the agreement between theory and experiment nontrivial. Overall, whether the SLC leads to an amplification or an attenuation depends on the spin Hamiltonian and the particular form of the SLC. Examples for the amplification by SLC in the HKGG' model are reported in the SI with different parameters. A systematic study is needed to predict which of these two scenarios can be expected in other systems.

The model including SLC achieves an agreement between theoretical and experimental results in overall sign, magnitude, and the general field dependence, in contrast to the pure magnon calculations. The remaining quantitative disagreement between the effective theoretical model and experiment could be caused by the presence of multiple domains close to the phase transition, which is not accounted for in our model, the restriction to one phonon band and one particular type of SLC, and the disregard of vibrational degrees of freedom of nonmagnetic ions. The deviation between the minima of $\kappa_{xy}(B_z)$ measured at 3 T and computed at 10 T may be attributed to extrinsic contributions to the THC, as indicated by the correlation between the measured minima of $\kappa_{xx}$ and $\kappa_{xy}$ at similar fields (compare Figures 2d,e and 3a). Hence, at lower fields, extrinsic contributions appear to be relevant for a better quantitative agreement, while at larger fields, due to the lack of a similar prominent correlation, their relevance might be limited. Therefore, the extrinsic contributions to the THC such as magnon-phonon scattering, magnon-magnon scattering, and scattering of phonons or magnons at (magnetic) impurities should be investigated in a more comprehensive quantitative theory.

An open question for NCTO is whether its ground state is of zigzag antiferromagnetic or of triple-Q nature. While several studies have argued in favor of triple-Q,[32,34,35] another reports inconsistent observations with the triple-Q ground state.[33] Our study shows that the zigzag antiferromagnetic ground state is compatible with THC measurements, however, we cannot conclusively rule out the possibility of a triple-Q ground state. Whether the triple-Q ground state is also compatible with our THC measurements needs to be addressed in the future.

Finally, our results, in particular the fact that magnon polarons and pure magnons can drive opposite heat currents of different magnitudes, demonstrate that the SLC may completely alter the low-temperature transport properties and overshadow predicted transport signatures of isolated quasiparticles, like topological magnons. Instead of transport signatures of isolated exotic spin excitations, a more unified approach that includes the hybridization with phonons is necessary for the interpretation of such transport experiments. To verify the importance of SLC in NCTO, but also more generally, an independent determination of the SLC by ab initio calculations or magnetoelastic experiments is required that should be combined with model calculations to quantify the impact on the THC. In short, our results call for a systematic analysis of the role of SLC in the THC.

**Methods**

**Single crystal growth.** The polycrystalline sample of NCTO was mixed with a flux of $Na_2O$ and $TeO_2$ in a molar ratio of 1:0.5:2 and gradually heated to 900 °C at 3 °C/minute in the air after grinding. The sample was retained at 900 °C for 30 hours and was cooled to a temperature of 500 °C at the rate of 3 °C/hour. The furnace was then shut down, allowing the sample to cool down to room temperature. The obtained NCTO single crystals are thin plates with hexagonal geometry. The orientation of the crystals was confirmed by X-ray Laue back diffraction measurement. The magnetic ordering temperatures shown by the magnetic susceptibility data (cf. SI) are consistent with the reported data.

**Heat transport measurements.** A thin-plate shaped crystal with size of $6.3 \times 2.17 \times 0.053$ mm$^3$ was used for heat transport measurements. The longitudinal thermal conductivity and thermal Hall conductivity were measured simultaneously by using the standard steady-state technique with "one heater, three thermometers". Heat current and magnetic field were applied along the *a* and *c* axes, respectively, as illustrated in Figure 2b. The longitudinal and transverse temperature gradients were measured by three in situ calibrated RuO$_2$ thermometers. The measurements were carried out in a $^3$He refrigerator equipped with a 14 T magnet. The longitudinal thermal conductivity $\kappa_{xx}$ and thermal Hall conductivity $\kappa_{xy}$ were obtained by the following formulas[43,42]

$$\omega_{xx} = \nabla_x T / j_q,$$
$$\omega_{xy} = \nabla_y^{\text{asym}} T / j_q,$$
$$\kappa_{xx} = \frac{\omega_{xx}}{\omega_{xx}^2 + \omega_{xy}^2},$$
$$\kappa_{xy} = -\frac{\omega_{xy}}{\omega_{xx}^2 + \omega_{xy}^2},$$

where $j_q$, $\nabla_x T$, $\omega_{xx}$, and $\omega_{xy}$ are the heat current, longitudinal temperature gradient, longitudinal thermal resistivity, and thermal Hall resistivity, respectively. In order to avoid the longitudinal response originating from the misalignment of two transverse contacts, the relationship $\nabla_y^{\text{asym}} T(B) = [\nabla_y T(B) - \nabla_y T(-B)]/2$ was used to remove the field-symmetric component. $\kappa_{xx}$ and $\kappa_{xy}$ were measured simultaneously at a fixed temperature while changing the magnetic field. In order to account for the possible magnetic hysteresis, at each temperature the sample was firstly zero-field cooled from 30 K (above $T_N$) to the target temperature, and the positive-field measurements were carried out with changing magnetic field from 0 to 14 T step by step; after the field reached 14 T, the sample was again zero-field cooled from 30 K and then the negative-field measurements were carried out with changing magnetic field from 0 to 14 T. As demonstrated in the SI, $\kappa_{xx}$ only shows a weak hysteresis. The thermometers (Cernox and RuO$_2$) mounted on the sample stage are pre-calibrated by using a capacitance sensor (Lakeshore Cryotronics) as the reference. The resolutions of the $\kappa_{xx}$ and $\kappa_{xy}$ measurements are typically better than 2 % and 10 % (with respect to the maximum absolute values of each field-dependent measurement max $|\kappa_{xx}(B)|$ and max $|\kappa_{xy}(B)|$), respectively.

**Theoretical model and transport calculations.** The Heisenberg-Kitaev-Gamma-Gamma' Hamiltonian reads

$$H = \frac{1}{2\hbar^2}\sum_{\langle ij\rangle_r} J_r \mathbf{S}_i \cdot \mathbf{S}_j + \frac{1}{2\hbar^2}\sum_{\langle ij\rangle}\left[K S_i^\gamma S_j^\gamma + \Gamma\left(S_i^\alpha S_j^\beta + S_i^\beta S_j^\alpha\right) + \Gamma'\left(S_i^\gamma S_j^\alpha + S_i^\gamma S_j^\beta + S_i^\alpha S_j^\gamma + S_i^\beta S_j^\gamma\right)\right]$$

($\hbar$ reduced Planck's constant). It encompasses Heisenberg exchange $J_r$ ($r = 1, 2, 3$) up to 3$^{\text{rd}}$ nearest neighbors $\langle ij\rangle_3$ and Kitaev ($K$), Gamma ($\Gamma$), Gamma' ($\Gamma'$) interactions between nearest neighbors $\langle ij\rangle$. The mutually orthogonal spin components $\alpha, \beta, \gamma = x, y, z$ depend on the bond between sites $i$ and $j$; they are expressed in a basis in which the [111] direction is parallel to the $c$ axis and all three axes are perpendicular to the honeycomb bond vectors. For each bond, $\gamma$ represents the bond-perpendicular axis, while $\alpha$ and $\beta$ are the components to the remaining two axes. The magnetic field $B$ enters via the Zeeman Hamiltonian

$$H_B = \frac{g\mu_B}{\hbar} \boldsymbol{B} \cdot \sum_i \boldsymbol{S}_i,$$

where $g = 2.3$ is the $g$-factor of NCTO[46] and $\mu_B$ is Bohr's magneton. Here, we are interested in out-of-plane fields $\boldsymbol{B} \parallel \boldsymbol{c}$.

In the following, the coordinate axes are changed such that the $z$ axis is pointing out-of-plane, i.e., $\hat{\boldsymbol{z}} \parallel \boldsymbol{c}$. We describe the out-of-plane lattice vibrations with a simple spring model

$$H_p = \sum_i \frac{(p_i^z)^2}{2M} + \frac{C}{4}\sum_{\langle ij \rangle}(u_i^z - u_j^z)^2$$

that is solved for the phonon eigenfrequencies and eigenvectors by diagonalizing the corresponding dynamical matrix (cf. SI). Spin-lattice coupling, which is derived from spin-orbit coupling, is achieved by

$$H_{\text{me}} = \frac{\tilde{\lambda}}{\hbar^2}\sum_i \sum_{\boldsymbol{\delta}} (\boldsymbol{S}_i \cdot \boldsymbol{\delta}) S_i^z (u_i^z - u_{i+\boldsymbol{\delta}}^z),$$

where the spin at site $i$ is projected onto the nearest-neighbor bond directions $\boldsymbol{\delta}$ and coupled to the $z$ displacements of the magnetic $Co^{2+}$ ions (cf. SI). A similar form of SLC was considered for ferromagnets and antiferromagnets with out-of-plane magnetic order.[25,26] We convert the interaction parameter $\tilde{\lambda}$ to an energy as $\lambda = \tilde{\lambda} d_{\text{nn}} \sqrt{\frac{\hbar M}{2C}}$. $C = 298.93 \frac{\text{meV}}{\text{Å}^2}$ is the elastic constant, $d_{\text{nn}} = 3.0361$ Å is the (in-plane) nearest-neighbor distance, and $M$ is the mass of the $Co^{2+}$ ions.

We then determine the classical ground state orientations $\hat{\boldsymbol{z}}_i$ of the spins of each sublattice $i = 1, 2, 3, 4$ and define orthogonal local axes $\hat{\boldsymbol{x}}_i$ and $\hat{\boldsymbol{y}}_i$. The spin operators are represented by bosonic operators using the Holstein-Primakoff transformation[47]

$$\frac{\boldsymbol{S}_i}{\hbar} = (S - a_i^\dagger a_i)\hat{\boldsymbol{z}}_i + \sqrt{\frac{S}{2}}(a_i^\dagger f_i \hat{\boldsymbol{e}}_i^+ + f_i a_i \hat{\boldsymbol{e}}_i^-),$$

where $\hat{\boldsymbol{e}}_i^\pm = \hat{\boldsymbol{x}}_i \pm i\hat{\boldsymbol{y}}_i$ and $f_i = \sqrt{1 - \frac{a_i^\dagger a_i}{2S}}$ is approximated as 1 (harmonic approximation). The bosonic Hamiltonian is expanded in terms of fluctuations about the ground state. After the diagonalization of the bilinear Hamiltonian, we obtain the four magnon states $|n\boldsymbol{k}\rangle$ with energies $\varepsilon_{n\boldsymbol{k}}$, where $n$ is the band index and $\boldsymbol{k}$ is the wave vector, that are employed to compute the Berry curvature[48] $\Omega_{n\boldsymbol{k}}$ using the method by Fukui *et al*.[49]

The linear-response expression for the intrinsic THC is reads[3,4]

$$\kappa_{xy} = -\frac{k_B^2 T}{\hbar V} \sum_{n\bm{k}} c_2[\rho(\varepsilon_{n\bm{k}})]\Omega_{n\bm{k}},$$

where $k_B$ and $V$ are the Boltzmann constant and sample's volume, respectively. Note that $V = Ad$, where $A$ is the area of the 2D sample and $d = c/2 = 5.5850\,\text{Å}$ is the inter-layer distance. The summands comprise $c_2(x) = (1+x)\left(\ln\frac{1+x}{x}\right)^2 - (\ln x)^2 - 2\text{Li}_2(-x)$ with the Spence's function $\text{Li}_2$, the Bose-Einstein distribution $\rho(\varepsilon) = \left[\exp\left(\frac{\varepsilon}{k_B T}\right) - 1\right]^{-1}$, and the Berry curvature $\Omega_{n\bm{k}}$.

Quantum and thermal spin fluctuations are calculated in linear spin wave theory by evaluating the expectation value $\langle S_i^z \rangle = \langle S - a_i^\dagger a_i \rangle$ at zero and nonzero temperatures, respectively. We rescale the magnon energies by $\Lambda = \langle S \rangle/S$,[50] which we approximate by 0.8, suggested by the weak dependence of $\langle S \rangle = \langle S_i^z \rangle$ on sublattice $i$, temperature and magnetic field (cf. SI). The superscript $z$ refers to the spin component along the ground spin orientation $\hat{\bm{z}}_i$.

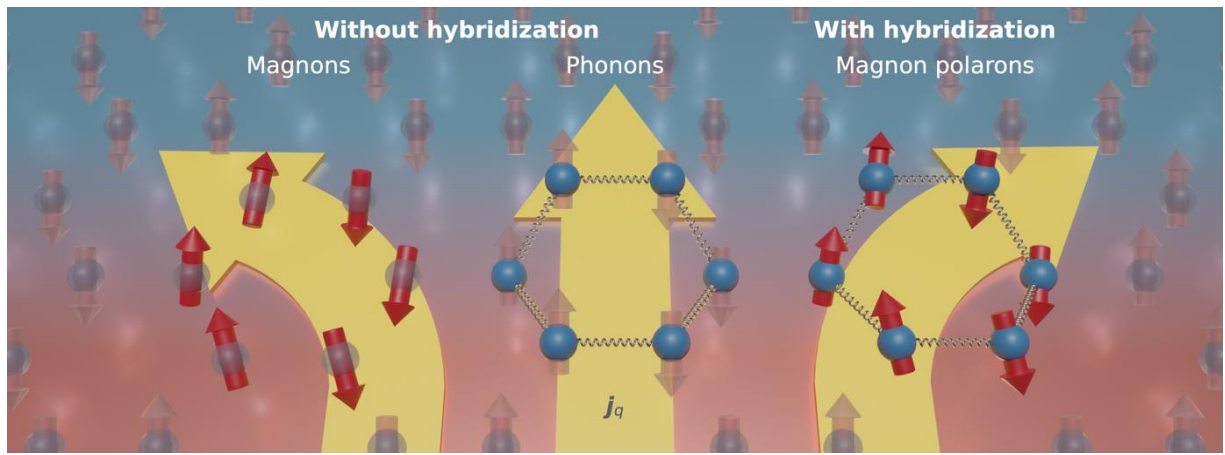

**Figure 1 | Qualitative visualization of our theoretical results.** Intrinsic thermal transport of distinct quasiparticles moving from hot to cold in a temperature gradient. Without hybridization, magnons contribute to the longitudinal and transversal transport, while phonons only contribute to the longitudinal one (in our approximation). With hybridization, magnons and phonons merge into magnon polarons and the transverse transport direction is reversed.

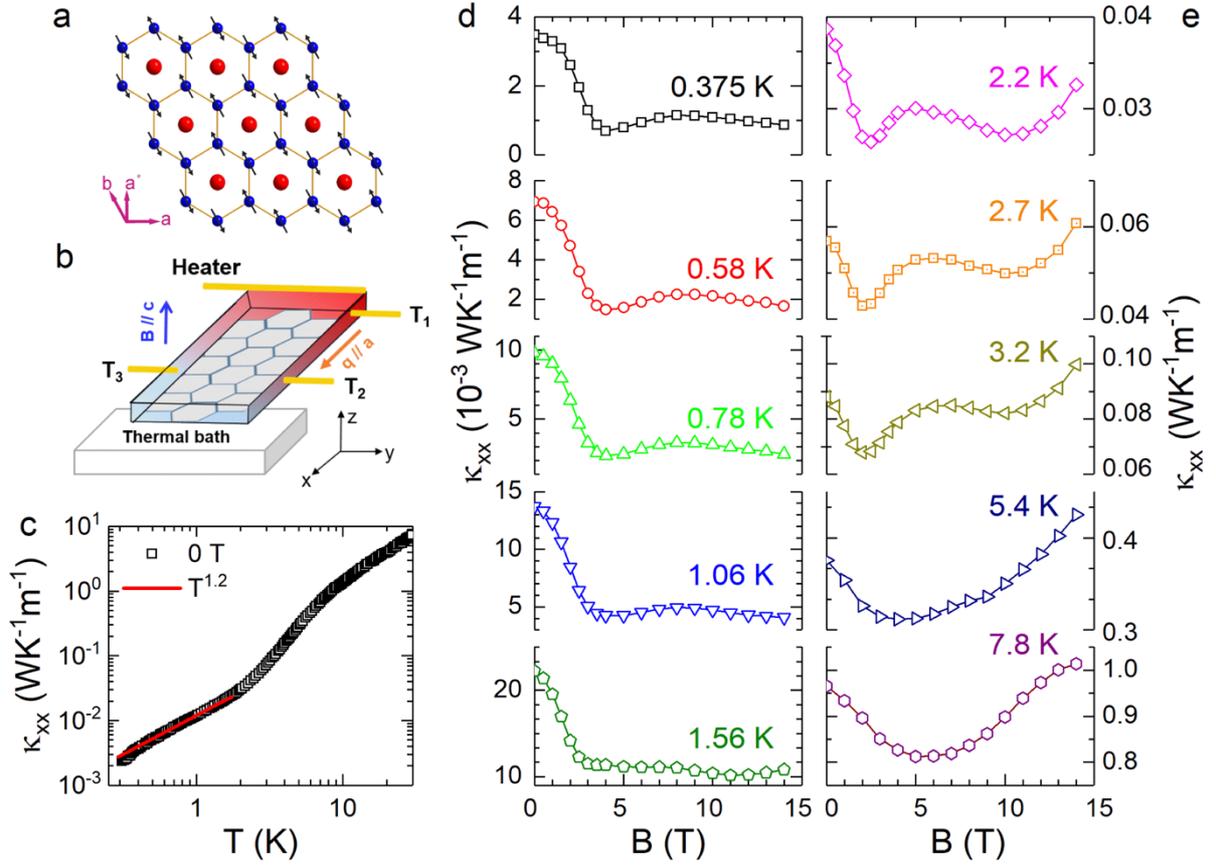

**Figure 2 | Longitudinal thermal heat conductivity $\kappa_{xx}$ of a Na$_2$Co$_2$TeO$_6$ single crystal. a,** Crystallographic spin structure of the $ab$ plane of NCTO in the AFM state. The honeycomb lattice consists of Co ions (blue spheres) in the zigzag AFM arrangements (indicated by arrows), and the Te ions (red spheres) are located at the center of each honeycomb. The $a^*$ axis is the in-plane direction perpendicular to the $a$ axis. **b,** Schematic of the experimental setup for the thermal Hall measurements. The heat current and the magnetic field are applied along the $a$ and $c$ axis, respectively. The longitudinal and transverse temperature gradients are determined by the difference between $T_1$ and $T_2$ and between $T_2$ and $T_3$, respectively. **c,** Temperature dependence of the longitudinal thermal conductivity $\kappa_{xx}$ at zero magnetic field. The zero-field data roughly display a $T^{1.2}$ behavior at very low temperatures, as the solid line indicates. **d, e,** Magnetic-field dependence of the thermal conductivity at various temperatures and with $\boldsymbol{B} \parallel \boldsymbol{c}$.

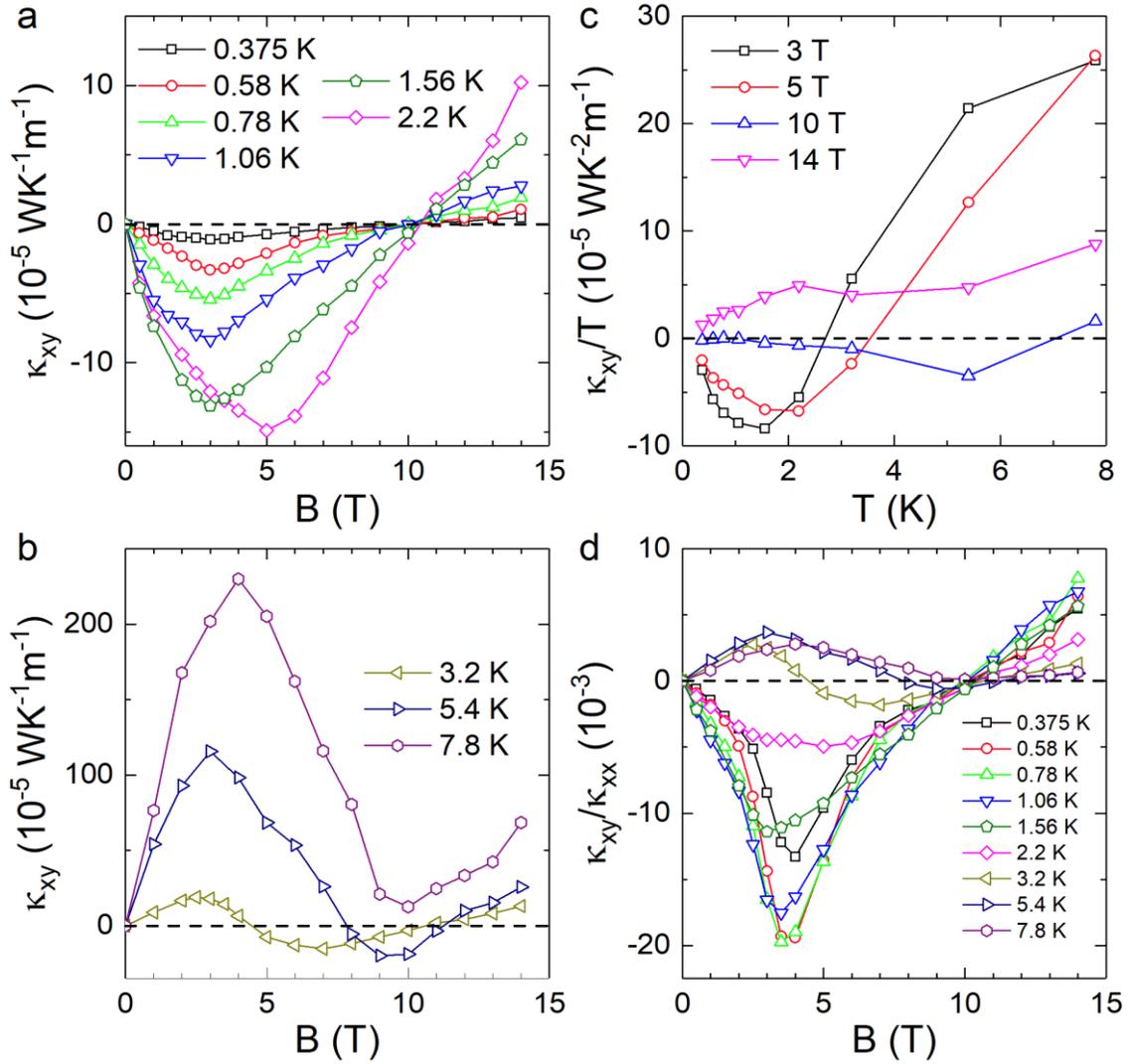

**Figure 3 | Thermal Hall conductivity of a Na$_2$Co$_2$TeO$_6$ single crystal. a, b,** Field dependence of thermal Hall conductivity $\kappa_{xy}$ for $\boldsymbol{B} \parallel \boldsymbol{c}$ at various temperatures. **c,** Temperature dependence of $\kappa_{xy}/T$ at selected magnetic fields. **d,** Magnetic field dependence of the thermal Hall angle $\kappa_{xy}/\kappa_{xx}$ at various temperatures.

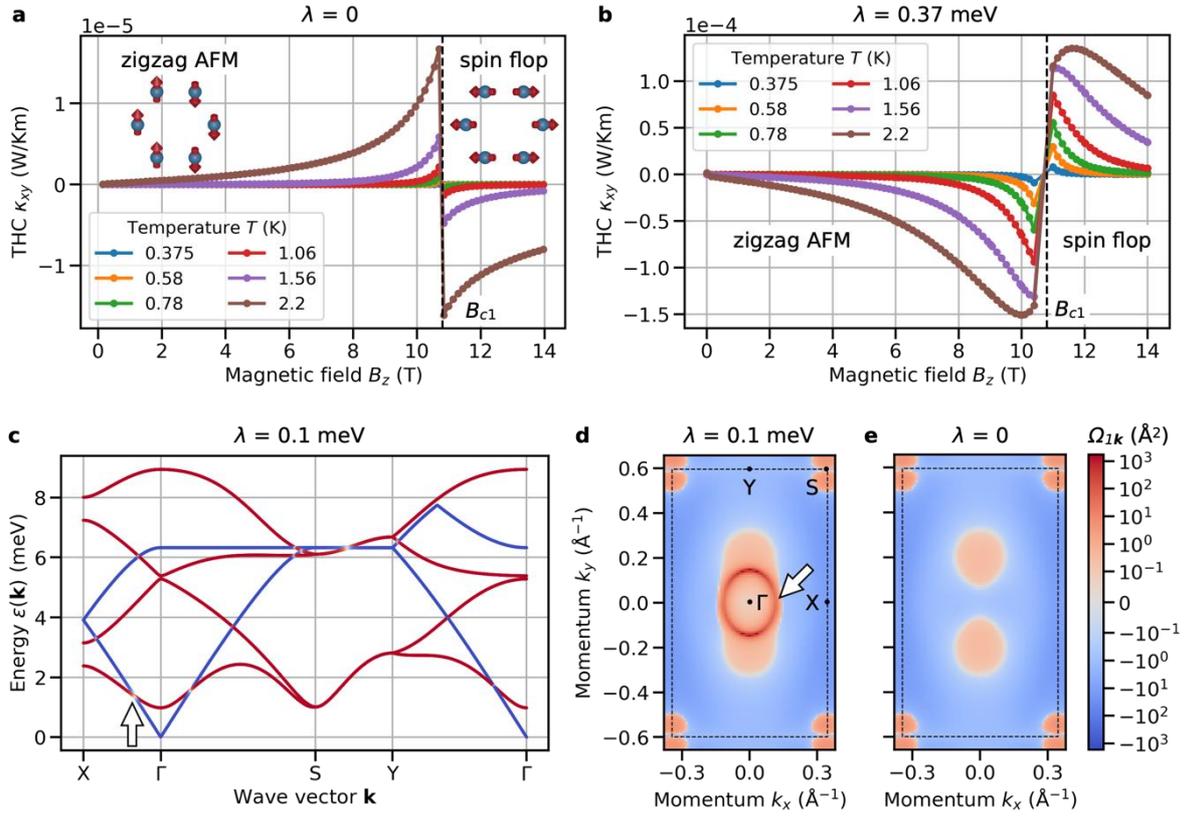

**Figure 4 | Model calculations. a, b,** Thermal Hall conductivity $\kappa_{xy}$ vs applied field $B_z$ (**a**) without ($\lambda = 0$ meV) and (**b**) with SLC ($\lambda = 0.37$ meV). Inset: magnetic ground state of Co$^{2+}$ ions in their two respective phases. **c,** Magnon-polaron spectrum $\varepsilon_{n\mathbf{k}}$ without fluctuations ($\Lambda = 1$) along a high-symmetry path. Red/white/blue colored bands indicate magnon/mixed/phonon character of the modes; the magnetic field is 5 T. **d, e,** Berry curvatures $\Omega_{n\mathbf{k}}$ of the lowest bands $n = 1$ (**d**) with SLC ($\lambda = 0.1$ meV) and (**e**) without SLC ($\lambda = 0$ meV). Dashed rectangles in **d, e** mark the first Brillouin zone. Other parameters are given in the text.

**Data availability**

The data that support the findings of this study are available from the corresponding author upon reasonable request.

**Acknowledgements**

We thank I. Kimchi for insightful discussions, and W. J. Chu and X. H. Zhou for their help with the experiments. This work was supported by the National Natural Science Foundation of China (Grant Nos. 12274388, 12174361, 12025408, 11904003, and 12274001) and the Nature Science Foundation of Anhui Province (Grant Nos. 1908085MA09, 2108085QA22, and 2208085MA09). The work at the University of Tennessee and Georgia Tech was supported by the U.S. Department of Energy under Award Nos. DE-SC-0020254 and DE-FG02-07ER46451. A. M., J. H., and I. M. acknowledge funding from Deutsche Forschungsgemeinschaft (DFG, German Research Foundation) - Project No. 504261060 and SFB TRR 227.

**Author contributions**

N. L., S. K. G., K. X., X. Y. Y., Y. Y. W., Y. S., and Q. J. L. performed heat transport measurements and analyzed the data with help from X. Z., X. F. S. and H. D. Z. Q. H., J. L., and H. D. Z. made the single-crystal samples. R. R. N., Y. J., J. F., Z. J., and A. M. performed the linear spin wave and magnon thermal Hall calculations. For the magnon-phonon model, R. R. N., with the help of A. M., J. H., and I. M., developed the theory and R. R. N. wrote the code. All authors discussed the results. R. R. N., Z. J., A. M., J. H., I. M., H. D. Z., and X. F. S. wrote the manuscript.

**Competing financial interests**

The authors declare no competing financial interests.

**Additional Information**

Correspondence and requests for materials should be addressed to I. M., H. D. Z., or X. F. S.